
\input phyzzx
\sequentialequations
\overfullrule=0pt
\tolerance=5000
\nopubblock
\twelvepoint

\REF\susskindfive{L. Susskind and J. Uglum,
Phys. Rev {\bf D50} (1994) 2700}

\REF\barbon{J. L. F. Barb\'{o}n and R. Emparan,
{\it On quantum black hole entropy and Newton coupling constant
renormalization},
hep-th/9502155},

\REF\demers{J-G Demers, R. Lafrance, and R. C. Myers,
{\it Black hole entropy without brick walls},
gr-qc/9503003}

\REF\hawkingtwo{G.W.Gibbons and S.W. Hawking,
Phys. Rev. {\bf D15} 2752 (1977).}

\REF\hawkingthree{S.W. Hawking,
Phys. Rev. {\bf D18} 1747 (1978).}

\REF\singer{H. P. McKean Jr. and I. M. Singer,
J. Diff. Geom. 4 (1967) 43.}

\REF\balian{R. Balian and C. Bloch,
Ann. Phys. {\bf 64} (1971) 271.}

\REF\fursaev{D. V. Fursaev and S. N. Solodukhin,
{\it On One-Loop Renormalisation of Black hole},
gr-qc-9412020}

\REF\solodukhin{S. N. Solodukhin,
{\it One-loop Renormalization of Black Hole Entropy Due to
Non-minimally Coupled Matter},
hep-th/9504022}

\REF\bombelli{L. Bombelli, R. Koul, J. Lee, and R. Sorkin,
Phys. Rev.  {\bf D34} (1986) 373.}

\REF\srednickione{M. Srednicki,
Phys. Rev. Lett. {\bf 71}, (1993) 666. }

\REF\hlw{ C. Holzhey, F. Larsen and F. Wilzcek,
Nucl. Phys. {\bf B424} (1994) 443}

\REF\ccfw{C.G.Callan and F. Wilczek,
Phys. Lett. {\bf B333} (1994) 55.}

\REF\kabat{D. Kabat and M.J.Strassler,
Phys. Lett. {\bf B329} (1994) 46 }

\REF\dowkerone{J.S.Dowker,
Class. Quant. Grav. 11 (1994) L55}

\REF\fermions{F. Larsen and F. Wilczek,
{\it Geometric Entropy, Wave Functionals, and Fermions},
PUPT 1480, IASSNS 94/51,hep-th/9408089}

\REF\kabattwo{D. Kabat,
{\it Black Hole entropy and entropy of entanglement},
hep-th/9503016}

\REF\teitelboimfour{M. Banados, C. Teitelboim, and J. Zanelli,
Phys. Rev. Lett. {\bf 72} (1994) 957}


\REF\cardy{J. L. Cardy,
{\it Conformal Invariance and Statistical Mechanics},
in {\it Fields, Strings, and Critical Phenomena}
Les Houches, Session XLIX, 1988, eds. E.Br\'{e}zin and J. Zinn-Justin.}

\REF\vassilevich{D. V. Vassilevich,
{\it QED on Curved Background and on Manifolds with Boundaries:
Unitarity versus Covariance},
gr-qc/9411036}

\REF\birrell{N. D. Birrell and P.C.W. Davies,
{\it Quantum Fields in Curved Space},
Cambridge University Press, Cambridge (1982)}

\REF\thooftone{G. 't Hooft,
Nucl. Phys. {\bf B256} (1985) 727}

\REF\adler{S. L. Adler,
Rev. Mod. Phys. {\bf 54} (1982) 729.}

\REF\gsw{M. B. Green, J. H. Schwarz, and E. Witten,
{\it Superstring Theory},
Cambridge University Press (1987)}

\REF\bekenstein{J. Bekenstein,
{\it Do we understand Black Hole Entropy?},
gr-qc/9409015 }

\REF\perthree{ For a first step in this direction see
P. Kraus and F. Wilczek, Nucl. Phys. {\bf B433} (1995) 403}

\line{\hfill }
\line{\hfill PUPT 1541, IASSNS 95/49}
\line{\hfill hep-th/9506066 }
\line{\hfill June 1995}

\titlepage
\title{Renormalization of Black Hole Entropy and
of the Gravitational Coupling Constant}

\author{Finn Larsen\foot{Research supported in part by a Danish National
Science Foundation Fellowship.\ \ \ larsen@puhep1.princeton.edu}}
\centerline{{\it Department of Physics }}
\centerline{{\it Joseph Henry Laboratories }}
\centerline{{\it Princeton University }}
\centerline{{\it Princeton, N.J. 08544 }}
\vskip .2cm
\author{Frank Wilczek\foot{Research supported in part by DOE grant
DE-FG02-90ER40542.~~~wilczek@iassns.bitnet}}
\vskip.2cm
\centerline{{\it School of Natural Sciences}}
\centerline{{\it Institute for Advanced Study}}
\centerline{{\it Olden Lane}}
\centerline{{\it Princeton, N.J. 08540}}

\endpage

\abstract{The quantum corrections
to black hole entropy, variously defined, suffer
quadratic divergences reminiscent of the ones found in the renormalization
of the gravitational coupling constant
(Newton constant). We consider the suggestion, due
to Susskind and Uglum,
that these divergences are proportional,
and attempt to clarify its
precise meaning.
We argue that if the black hole entropy is identified using a Euclidean
formulation, including the necessary surface term as proposed by
Gibbons and Hawking, then the proportionality is,
up to small
identifiable corrections, a
fairly immediate consequence of basic principles
-- a low-energy theorem.
Thus in this framework renormalizing the
Newton constant renders the entropy finite, and
equal, in the limit of large mass, to its semiclassical value.
As a partial check on our formal arguments
we compare the one loop determinants,
calculated using heat kernel regularization.
An alternative definition of black hole entropy
relates it to behavior at conical singularities
in two dimensions, and
thus to a suitable definition of geometric entropy.
A definition of geometric
entropy, natural from the point of view of heat kernel
regularization,  permits the same renormalization,
but it does not yield an intrinsically positive
quantity.
The possibility, for scalar fields,
of
non-minimal coupling to background curvature
allows one to consider test the framework in
a continuous family of theories, and crucially involves a
subtle sensitivity of geometric entropy to curved space couplings.
Fermions and gauge fields are considered as well.  Their functional
determinants
are closely related to the determinants for non-minimally coupled scalar
fields with specific values for the curvature coupling, and pose no
further
difficulties.}

\endpage

\chapter{Introduction}

It has been proposed that
the divergences of the entropy in black hole thermodynamics have
the same origin as, and indeed are proportional to,
the divergences of the gravitational coupling
constant in na\"{\i}ve perturbative quantum gravity [\susskindfive].
The possibility of such a connection
is certainly appealing, but
several objections have been raised to it [\susskindfive --\demers ].
For one thing, the divergences arising in renormalization of $G$
are certainly sensitive to non-minimal couplings of the matter fields
to curvature,
whereas the relevant entropy can be identified in flat space.  Also,
the divergent renormalizations, at one loop,
can have either sign depending on the spin and
curvature
coupling of the field involved, whereas the entropy would appear to
be intrinsically positive by definition.
Moreover,
since both sides of the proposed equality are infinite, and the precise
definition of one side (the black hole entropy) is notoriously
controversial, clearly
some non-trivial questions of interpretation are involved.
In this paper we shall propose an interpretation in which the claim
is both precise and true, as a low-energy theorem.  We shall also discuss
the tensions that arise in other interpretations, and show that at
least some of these -- specifically, the two mentioned above --
are less severe than appears at first sight.

We will first consider the definition of
black hole entropy proposed by Gibbons and Hawking [\hawkingtwo--
\hawkingthree ],
within their Euclidean approach to quantum
gravity.
If we accept that framework for considering
black hole entropy, then this entropy
arises from a surface term in the effective
action, whose coefficient is related in a precise
numerical fashion
to the bulk
Einstein-Hilbert term.  The coefficient of the Einstein-Hilbert
term, of course, in turn
defines the observed
Newton constant $G$.
Thus one obtains, in the limit of large black
holes,
a low-energy theorem for the black hole entropy,
expressing
it directly in terms of the fully renormalized Newton constant.
This result relies only on the  structure of the action, so it is
valid
upon the rather mild assumption that the effective quantum action has
the
same structure as the classical one.
In this regard, note that to treat large black holes
in the Euclidean formalism
one need only consider smooth manifolds of uniformly small curvature.

Unfortunately
these arguments are of course purely formal, since there are
serious problems with the ultraviolet behavior of the underlying
theory, and all the quantities involved are infinite unless regulated.
Within the Euclidean framework the divergences
in the entropy and in the quantum corrections to Newton's constant
have a common origin in local vacuum polarization.
Heat kernel methods provide an appropriate
way to regulate such divergences for the one-loop contribution
of matter fields,
while maintaining symmetry and locality [\singer --\solodukhin].
Using
this
regularization, we calculate
the leading cutoff dependence explicitly
in a uniform manner for various
spins and statistics (and for minimal or non-minimal coupling).

The Gibbons--Hawking definition of black hole
entropy does not on the face of it
offer a satisfactory
understanding of this
entropy based on the same principles as conventional
definitions of entropy in statistical physics.  Thus it is
not superfluous to consider
other definitions of entropy that are closer in spirit
to the
conventional definitions.  Recently geometric entropy, which has
considerable intrinsic interest,
has been extensively
discussed in this context.
[\bombelli--\fermions ]
We show that the
geometric entropy relevant to black hole physics features
corrections, in the form of winding modes, that are non-local from the
point of view of particle paths.
These are responsible for divergences which,
on the face of it, appear to
have a very different origin than those arising in
renormalizing the Newton constant.
We find that, with natural definitions,
the divergences in the geometric entropy and
the gravitational coupling nevertheless coincide.
This result emerges for reasons that are, at least to us, rather
less transparent than in the alternative (by no means
obviously equivalent) Euclidean framework.
It is particularly interesting, for reasons mentioned already, to consider
the effects of non-minimal curvature couplings.
We show that such non-minimal couplings reflect themselves even in
flat space, where they dictate the form
of the local energy-momentum tensor.  Specifically, they control
total divergence terms, which do not affect the integrated
energy-momentum but
do affect the regulation of modes on a half-space, and thereby sneak
into the calculation of the properly regulated geometric entropy.

While we were in the final stages of preparing this paper we received
an important paper by Kabat [\kabattwo],
which overlaps in part with our discussion
of geometric entropy, while featuring quite different techniques and
emphases.  We shall make some more detailed comparisons below.


\chapter{Low-Energy Theorem for Gibbons-Hawking Entropy}

There are several definitions of the
black hole entropy in common use, whose equivalence is not
manifest.
In this section we consider the entropy defined
using a Euclidean path
integral for quantum gravity, following Gibbons and
Hawking [\hawkingtwo -\hawkingthree].  This definition came very early
historically, but has not been so prominent in the recent literature.
Since
it is the definition which makes the non-renormalization almost obvious,
we will give an elementary, self-contained
presentation of it in this Section.
We also take this opportunity to fix conventions and notations.

Within ordinary quantum field theories one
can define the canonical ensemble, fixing the temperature $T$ of the
system, by a Euclidean path integral over all configurations subject
to the constraint that they are periodic (antiperiodic for fermions)
in imaginary time
$\beta\equiv 1/T$.  The ``imaginary time'' is a real variable,
of course; the integrals are defined by analytic continuation, not
substitution, from the real-time integrals.
This prescription is essentially the same as the
Kubo-Martin-Schwinger boundary condition, which can be derived from basic
principles of axiomatic field theory.  For several
reasons these basic principles do not apply
cleanly to general relativistic systems, so there is a leap of faith
involved in the use of the Euclidean path integral for gravity.
In this chapter we shall take the leap,
and consider the path integral over
Euclidean space-times as defining the partition function.
(The Euclidean formalism does not give rise to any obviously
problematic
results for the kinds of calculations we do in this paper, involving
the quantum mechanics of matter fields in simple curved spaces.  However
if we attempted to calculate the vacuum polarization due
to gravitons -- the spin-two functional determinant -- we would meet
with
the notorious difficulties associated with the non-positivity of the
action for the conformal factor [\hawkingthree ].)

It is very important, in attempting to access the thermodynamics of a
black hole, to consider a finite volume.  Indeed the black hole
contribution
to thermodynamic functions is finite rather than extensive in the
volume, and it will always be swamped, in a large enough volume, by
the contribution of the ambient thermal bath.
In setting up the finite volume integral,
one must complement the
standard Einstein-Hilbert action with
a surface term.
This arises because higher derivatives occur in the
Einstein-Hilbert action, whose presence would
invalidate the variational principle
in its usual form.  Fortunately the offending terms can be isolated,
after an
integration by parts, and subtracted off as a boundary
term,  thus yielding
an action amenable to conventional path integral treatment.
The partition function then becomes
$$
Z=\int {\cal D}g~e^{-{1\over{\hbar}}L}~~;~~
L=-{1\over 16\pi G}(\int dV R - 2\int d\sigma K)
\eqn\partfnct
$$
where $K$ denotes the extrinsic curvature.
This is to be evaluated as a function of $\beta$ and the
geometry of the bounding surface.

The semiclassical approximation is implemented by evaluating the action on
some classical solution. In the zero angular momentum
vacuum sector, the unique
solutions are given by the Euclidean Schwarzschild metrics
$$
ds^2 = (1-{2MG\over r})dt^2 + (1-{2MG\over r})^{-1} dr^2 + r^2 d\Omega^2~.
\eqn\eucschw
$$
Introducing the coordinate ${\bar r}=\sqrt{8MG(r-2MG)}$,
one finds that the metric close to $r=2MG$
takes the form
$$
ds^2 = {{\bar r}^2\over 16M^2 G^2} dt^2 + d{\bar r}^2 + (2MG)^2 d\Omega^2~.
\eqn\metsing
$$
The equations of motion require the scalar
curvature to vanish everywhere.
In the preceding form of the metric it is
apparent that there is
a conical singularity at the ``horizon'' $\bar r = 0$ unless
$t$ is interpreted as an angular variable with period $8\pi MG$.
Since $R=0$ for a solution of the equations of motion, the value of the
action is entirely determined by the surface term.   We are interested in
the contribution of the non-trivial black hole topologies relative to the
flat space topologies, so we normalize
the path integral by a flat space solution with
the same interior geometry on the boundary surface.  Thus
we obtain
$$
{\rm ln}Z = -{1\over 8\pi G {\hbar}} \int d\sigma [K]
= - {\beta^2\over 16\pi G \hbar} ~~;~~~[K]= K^{BH}-K^{\rm vac}
$$
The second equality require a small calculation using
$\int d\sigma K = {\partial\over\partial {\hat n}} \int d\sigma$, which
we sketch in the following paragraph.
Here $\hat n$ is the inward pointing unit normal.

Choose the bounding surface to be $r=r_0$.  We demand that at this
surface the geometry be the product of a circle of length $\beta$ and
a sphere of area $4\pi r_0^2$.  For the Euclidean Schwarzschild solution,
this fixes $M$ in terms of $\beta$ according to
$$
(1-{2MG\over r_0})^{1/2} 8\pi MG ~=~ \hbar \beta~.
$$
The area of the 2-sphere as a function of $r$ is simply $4\pi r^2$ while
the length of the imaginary time circle is
$(1-{2MG\over r})^{1/2}  (1-{2MG\over r_0})^{-1/2} \beta $.  To calculate
$K^{BH}$ we
must take the derivative with respect to the unit vector,
{\it i.e}. $-(1-{2MG\over r})^{1/2} {\partial\over \partial r} $ of the
product of these factors, evaluated at $r= r_0$.
For flat space one easily finds $K^{\rm vac} = 8\pi \beta r_0$.
In the difference the term which grows with $r_0$ cancels, and the
term independent of $r_0$ gives the quoted result for the free energy.
Higher order terms involving $\beta/ r_0$ are neglected.

Now, from the standard thermodynamic formula
$$
S = -\beta^2 {\partial\over\partial\beta} {1\over\beta} {\rm ln}Z
$$
we obtain
$$
S = \hbar {\beta^2 \over 16 \pi G}~.
\eqn\entropybeta
$$
Then finally using the relationship between $\beta$ and $M$, we
arrive at the celebrated result
$$
S ~ = ~{4\pi GM^2 \over \hbar } ~=~ {1\over 4G\hbar} A~,
\eqn\entropyMA
$$
where $A= 4\pi (2GM)^2 $ is the area of the event horizon for a black hole
of mass $M$.

A non-renormalization theorem for black hole entropy, as
defined operationally following
the Gibbons-Hawking procedure outlined above, is an immediate corollary
of the structure of the calculation, given minimal assumptions regarding
the locality and general covariance of the effective Lagrangian.  Indeed,
the effective action will contain a term of the Einstein-Hilbert form,
which (in the absence of cosmological term) is the operator of lowest
mass dimension and dominates the long-distance
behavior for weak fields.  The coefficient of this term
for nearly flat space defines the renormalized Newton's
constant.   The surface term at large distances,
which as we have seen is responsible for the
entropy,  is uniquely determined -- including its numerical coefficient --
from the bulk term.  The asymptotic form of the metric at
infinity, up to the order in $1/r$
used in the calculation, is again determined in terms of the
coefficient of the Einstein-Hilbert term.
Thus, assuming only the validity of a general
covariant, local Lagrangian description of the dynamics at weak fields,
we arrive at \entropybeta\ in a form involving only renormalized
quantities.  Then from the thermodynamic formula
$M = E = F + S/\beta$ we find $\beta = 8\pi MG$, again in terms
of renormalized quantities, and thereby the first equality in
\entropyMA .   These results are low-energy theorems, in the sense familiar
from quantum field theory.


A minor but interesting point is that the Newton's constant is not quite
the empty-space Newston's constant, but rather that appropriate to
temperature $\beta^{-1}$.  These may differ by finite quantities, as
one integrates through mass thresholds or (taking into consideration
{\it e.g}.
$|\phi|^2 R$ terms, where $\phi$ is a Higgs field)  condensation scales.


The discussion so far has been extremely formal, in the sense that
neither side of the claimed equality is properly defined,
in view of various convergence problems in the quantum theory.  To some
extent this limitation is unavoidable, since no satisfactory
overarching theory of quantum gravity
is currently available in usable form.    Nevertheless some partial
calculations can be done as a consistency check.  Specifically, we can use
a heat kernel regularization of one-loop diagrams involving various
quantum fields, to get explicit results for the renormalization constants.
Since this regulator is local and general covariant, it embodies
the conditions for the non-renormalization theorem.

For simplicity, let us consider at this point
the contribution of
a minimally coupled scalar field.
The one-loop effective action $W$ is defined through
$$
e^{-W}=\int {\cal D}\phi~e^{-{1\over 8\pi}\int\phi (-\Delta+m^2)\phi}
=[{\rm det}(-\Delta+m^2)]^{-{1\over 2}}
\eqn\bosoneff
$$
We define the heat kernel
$$
D(\tau)={\rm Tr}e^{-\tau\Lambda}=\sum_i e^{-\tau\lambda_i}
$$
where $\lambda_i$ are the eigenvalues of $\Lambda=-\Delta + m^2$.
A short calculation shows
$$
W={1\over 2}{\rm ln}{\rm det}\Lambda={1\over 2}\sum_i {\rm ln}\lambda_i
=-{1\over 2}\int_{\epsilon^2}^\infty
d\tau {D(\tau)\over\tau}
\eqn\effaction
$$
The integral over $\tau$ does not converge for small $\tau$, so
we replace the lower limit with a non-zero cut-off $\epsilon^2$
and obtain a well--defined expression.
The leading divergence is related to the
short time behaviour of the heat kernel, which is independent
of mass. This problem is treated
abundantly in the
literature [\singer], so we only recall the main features of the calculation.
The heat kernel is given by
$$
D(\tau)=\int dx~G(x,x,\tau)
$$
where
the Green's function $G$ satisfies the differential equation
$$
({\partial\over\partial\tau}-\Delta_x)G(x,x^{\prime},\tau)=0~~;~~~
G(x,x^{\prime},0)=\delta(x-x^\prime)
$$
In flat space
$$
G_0(x,x^\prime,\tau) = ({1\over 4\pi\tau})^{d\over 2}
e^{-{1\over 4\tau}(x-x^\prime)^2}
$$
but in general we must expand in the Laplacian in local coordinates
and expand for small curvatures.
The result is [\balian]
$$
D(\tau)=({1\over 4\pi\tau})^{d\over 2}[\int dV
\pm {\sqrt{\pi\tau}\over 2}\int d\sigma
+{\tau\over 6}(\int dV R-2\int d\sigma K) + {\cal O}(\tau^{3\over 2})]
\eqn\smallcurv
$$
with Dirichlet and Neumann boundary conditions
corresponding to the upper and lower sign, respectively.

Integrating over $\tau$ we find the effective action for the
scalar field. The first two terms in the heat kernel can be
interpreted as a bulk cosmological constant and surface cosmological
constant, respectively. They contribute neither to the gravitational
coupling nor to the entropy.  They cancel in the latter, because the
{\it intrinsic\/} geometry of the Euclidean Schwarzschild bounding surface
is the same as that of the flat space boundary, whose action is subtracted
from it.
The subsequent two terms translate directly into
renormalisation of gravitational coupling and entropy, respectively.
Explicitly, defining the regularized Newton's constant
$$
{1\over 16\pi } {1\over G_{\rm ren}\hbar } ~=~
{1\over 16 \pi} ({1 \over G_{\rm bare} \hbar} + {a_G \over \epsilon^2} )
$$
and similarly the regularized entropy, one finds
$$
a_G=a_S={1\over 12\pi} {1\over\epsilon^2}~.
$$
A  mass for the scalar field does not change the leading divergence,
but contributes a logarithmic divergence and finite terms.
All such terms are equal for the coupling constant and the entropy.
Hence, in conclusion, we find the renormalized
Gibbons-Hawking entropy
$S={G_{\rm ren}M^2 \over \hbar}$, even after quantum corrections.

The non-renormalization of the Gibbons-Hawking entropy follows directly
from the explicit, known
form of the heat kernel, which only contains the Ricci scalar and
the extrinsic curvature of the boundary in the same combination
$\int dV R -2\int d\sigma K$ as in the tree level action.
This is as we anticipated above on very general grounds.  At the risk of
belaboring the point, let us restate the general argument in a
different language
more adapted to the spirit of the explicit calculation.
Upon making variations in the metric, no boundary term
remains in the combination $\int dV R -2\int d\sigma K$.
Physically, this is precisely
the condition that the energy-momentum tensor
contains no boundary term.
While physical membranes can be endowed
with surface tension, the boundary that we consider here
is a freely movable mathematical abstraction, and it had better not have
such a tension.
To insure the validity of the
variational principle, Lagrangeans on manifolds with
boundary are constructed such that
there is no energy--momentum on the boundary.
It should therefore come as no
surprise that, even after integrating the scalar field out,
no boundary energy--momentum is found. This argument appears to be
very general, applying for example to fields of any spin.
Moreover, it works entirely in the framework of smooth manifolds, and
weak curvature.
We assume only that variations in the metric can be
performed either
before or after the integration over fields, with concordant results.
A failure of this assumption, would imply that local Lorentz
invariance suffers an anomaly.

\chapter{Geometric Entropy and Cone Geometry}

\section{Generalities, and the Minimally Coupled Scalar}

In considering other possible definitions of the black hole
entropy, an important distinction must be drawn.
For on--shell definitions, which consider only regular geometries,
the preceding arguments apply.
An example of such a definition is the microcanonical
version (fix $M$, not $\beta$ ) of the Gibbons--Hawking entropy
[\teitelboimfour].
We implicitly relied on the equivalence of different
on--shell definitions, which
is a purely formal result, in the thermodynamic manipulations of the
previous section.
By contrast, in off--shell definitions one allows
$\beta$ and $M$ to vary independently.  As we have
discussed,  this will generally
introduce, in the Euclidean formulation,
a conical singularity at the horizon.
We must reconsider the preceding arguments,
for geometries of this kind.

Geometric entropy [\bombelli --\fermions] is an example
of an entropy implicitly defined off-shell.
It is explicitly defined in terms of microstates, as
$S=-{\rm tr}\rho{\rm ln}\rho$, where $\rho$ is the density matrix
obtained by tracing over some region of space.
In the limit of
interest for very large black holes the curvature is small
at the event horizon and the angular variables
decouple, so we are led to consider the trace over a half space.
In that case,
the geometric entropy is conveniently expressed
using the replica--trick,
$$
S_{\rm geom}= (1-n{d\over dn})_{n=1} {\rm ln}Z(n)
$$
where $Z(n)$ is the unnormalized
partition function of the field theory in question,
as calculated on the disc covered $n$ times.
As $n$ is required in the neighborhood of flat space,
$n=1$, $Z(n)$ can be thought of as the partition function on
a weakly singular
cone.


The conical singularity prevents us from using the small curvature
expansion, but
we can still use the functional determinant \effaction\
to calculate the partition function using the heat kernel.
On a product space the heat kernel is equal to product of
the heat kernels of each of the two spaces. The heat kernel on the
tranverse (angular) space is ${A\over 4\pi\tau}$ and we
are left to evaluate the heat kernel of the cone in $2$ dimensions.
It is conformally equivalent to the plane
so we can use techniques from conformal field theory, or others,
to
calculate the heat kernel exactly [\hlw, \dowkerone]. It is
$$
D(\tau) = {A\over 4\pi\tau}~{1\over 12} ({1\over n}-n )~.
\eqn\twodkernel
$$
In this expression we have retained
the transverse dimensions and
omitted terms proportional to the spacetime volume.

To find the geometric entropy from the effective action, we must
apply the operator $(1- n{\partial\over\partial n})$.
The result is
$$
S_{\rm geom} = {A\over 4}~{1\over 12\pi}~{1\over\epsilon^2}~.
\eqn\centres
$$
This result is precisely such that if we interpret it as the first
quantum
correction to the
classical entropy for a large black hole, we have
the non-renormalization
$$
S_{\rm total}=S^{\rm classical}+S^{\rm quantum}_{\rm geom}
= {A\over 4}~({1\over G_{\rm bare}{\hbar}} + {1\over 12\pi}~{1\over\epsilon^2})
= {A\over 4G_{\rm ren} {\hbar}}~,
\eqn\conenonren
$$
as for the Gibbons--Hawking entropy.


\section{Role of Winding Paths}

It is instructive
compare this treatment of the cone
partition function with what might be inferred from the
small curvature expansion of the heat kernel, \smallcurv .
On the cone there are no boundaries,
but the conical singularity contributes a
pointlike, delta-function curvature.  Let us
consider approaching this situation by first smoothing the singularity,
then taking the limit.
The Gauss-Bonnet theorem in two dimensions is
$$
{1\over 4\pi} (\int R - 2\int K ) = \chi = 2 - 2h - b
\eqn\gaussbonn
$$
where $h$ denotes the number of handles and $b$ the number of boundaries.
It gives us the relevant integral over curvature, regardless
of the details of the smoothing procedure, and we find
$$
D(\tau)\simeq {A\over 4\pi\tau}~{1\over 6}(1-n)~.
\eqn\smoothD
$$
Thus the small
curvature expansion does not correctly capture the heat kernel in a conical
background.
At large $n$, it differs from the exact result
(by a factor of 2), and at small $n$,
it completely misses the term including an inverse power of
$n$, which plays a crucial role in the evaluation of the
entropy.  Indeed
the term linear in $n$, which in smooth backgrounds is responsible
for coupling constant renormalization, does not
contribute to the entropy at all!  In physical terms,
this may be interpreted as a consequence of the
fact that local vacuum polarisation
affects only the normalization of the density matrix, but not
the entropy of entanglement.

The weak-curvature expansion of the heat kernel is derived from
purely local considerations, using Riemann normal coordinates [\balian ].
In the language of the underlying diffusion process, it does not take into
account paths that go around the cone.
We can roughly indicate the effect
of  these winding paths using the following procedure.
Consider the cone as the union
of infinitesmial slices with width $d{\bar r}$.
Each slice is a product manifold
of a small linear direction, and a periodic variable.
At fixed ${\bar r}$, the
coordinate $\theta=2\pi{\bar r}{t\over 8\pi MG}$ is periodic with period
$2\pi{\bar r}n$. The appropriate Greens function is
$$
G_0 = {d{\bar r}\over 4\pi\tau}~
\sum_{k=-\infty}^\infty e^{-{1\over 4\tau}
(\theta-\theta^\prime+2\pi{\bar r}nk)^2}
$$
The terms with non-zero $k$ represent field configurations
that wind around the cone. These terms are
non--perturbative in $\tau$ and were, correctly, ignored
in the small curvature expansion. They are not
inconsequential, however.
We take $\theta=\theta^\prime$ and integrate over space, to find the
heat kernel. It is
$$
D(\tau)= {A\over (4\pi\tau)^2}\int d{\bar r}(2\pi{\bar r}n)
\sum_{k\neq 0}~e^{-{1\over 4\tau}(2\pi{\bar r}nk)^2}
= {A\over 4\pi\tau}~{1\over 12} {1\over n}~.
$$
This reproduces the first term of the exact result.
A more refined version of this calculation, using the exact
kernel in polar coordinates, can be used to obtain \twodkernel\
in its entirety.

{}From this perspective, then, the divergence in geometric
entropy appears completely unrelated to perturbative
coupling constant renormalization.  Nevertheless, as we have seen,
the numerical evaluations agree.  One might rationalize this to
some extent by noting that the entropy calculation requires the
partition function only very near to flat space, where the smoothing
procedure is least drastic and the distinction between winding around
a weak pointlike curvature singularity and passing through a region
of high curvature is least distinct. This line of argument is
succesfully pursued in [\fursaev ].

\section{Non-minimal Couplings}

In general, the Lagrangean of a massless scalar field is
$$
{\cal L} = {1\over 8\pi}\phi (-\Delta + \xi R) \phi
\eqn\mzerol
$$
So far we have assumed minimal coupling $\xi = 0 $ for simplicity, but
in general we should include all dimension four operators on an equal
footing.

{}From the perturbative expansion of the heat kernel
$$
D(\tau) = ({1\over 4\pi\tau})^{d\over 2}
[\int dV +\tau ({1\over 6}-\xi ) \int dV R ]
\eqn\hkpert
$$
on a manifold without any boundary, we find the renormalized gravitational
coupling
$$
{1\over G_{\rm ren}\hbar} = {1\over G_{\rm bare}\hbar} + {1\over 2\pi} ({1\over
6}-\xi)
{1\over\epsilon^2}
\eqn\nmnewt
$$
The heat kernel on the cone with $n={\beta\over 8\pi MG}$ is
$$
D(\tau ) = {1\over 12} ({1\over n}- n )
- \xi (1 - n )
\eqn\hkcone
$$
Since  ${1\over 4\pi}\int R = 1-n$, the second term is the
perturbative result, which is expected to apply.
Indeed, write
$$
D(\tau) = {\rm Tr} e^{-\tau(-\Delta+\xi R)} \simeq
{\rm Tr} e^{\tau\Delta}[1+(e^{-\tau\xi R}-1)]
$$
The $\xi$-dependent term is explicitly suppressed by one power of $\tau$
relative to the leading term. For the leading term, the most
singular term (the volume term) is given accurately by its
perturbative form,
the winding-modes affecting only the subsequent order.
Similarly, for the $\xi$-dependent term,
it is sufficient to use
perturbative modes to evaluate this term to the leading order.
A rigorous calculation with the same result will be presented later
in this section.

Now, from the heat kernel on the cone we find the appropriate
effective action in four dimensions, which in turn leads
to the geometric entropy
$$
S_{\rm geom} =
{A\over 4} {1\over 2\pi} ({1\over 6} - \xi) {1\over\epsilon^2}
\eqn\nmgeo
$$
Hence, for the coupling constant and the entropy alike,
the quadratic divergences are proportional to ${1\over 6}-\xi$ and
otherwise independent of $\xi$. Therefore our conclusions from the previous
sections remain in the case of non-minimal coupling.
This result, simple as it appears, is somewhat surprising. The geometric
entropy, defined as $S=-{\rm tr}\rho{\rm ln}\rho$, is positive
definite for any finite matrix $\rho$. As we see now, the proper definition
of the formal expression does not in general lead to a positive definite
quantity.

The coupling to curvature enters the present calculation of
heat-kernel calculation of geometric
entropy because of the singular curvature on the cone. However,
geometric entropy is intrinsically
defined in a flat background where, one might think,
the coupling to background curvature is inconsequential.
What is going on here?

A related phenomenon, reviewed later in this section,
occurs in two dimensional conformal field
theory.  In that context,
addition of a term $\gamma R\phi$ to the Lagrangean changes
the energy-momentum tensor -- even in the limit of flat space
-- without destroying conformal invariance. The
central charge is changed to $c=1+12\gamma^2$.
The geometric entropy is proportional
to the central charge so, even in flat space, the
coupling to background curvature affects the result for
geometric entropy.

The physics behind these somewhat paradoxical results can be
understood qualitatively  as
follows.  The geometric entropy for a sharply sliced
half-space diverges, and only regulated forms of it can appear
in physical results.  If we wish to regulate it in a way that is
both local and covariant, it is more or less inevitable that we
must damp the contributions of high-frequency modes.  This also
corresponds to the realistic circumstance that the entropy associated
with arbitrarily high-frequency modes is not easily accessible to
observation.  Now to identify the high-frequency modes, we must know
the energy-momentum tensor.  Indeed, it is important that we know the
{\it local\/} energy-momentum tensor, so that we can identify these
modes near the boundary (where the divergences
arise).  The local energy-momentum tensor is
ambiguous up to a total divergence, if we merely demand that its
integral yield the conserved quantities.  However in extending the
theory in curved space, and demanding covariance, we must
remove this
ambiguity: the true local energy-momentum tensor can be identified by
varying the action with respect to the metric, according to
$$
 T_{\mu \nu} (x) ~=~ - {4\pi\over\sqrt g}
{\delta \over \delta g^{\mu \nu} }{\int {\cal L}\sqrt g}~.
\eqn\locem
$$
Following this procedure, one arrives at a
regulator which implicitly depends upon how one
extends the
theory into  curved space-time.  Thus even
apparently flat-space quantities, such as geometric entropy, that need
regulation can become connected -- through the demands of locality and
covariance -- to parameters governing
the behavior of the theory in curved space-time.


Now let us  consider concretely
the effect of the coupling $\xi R\phi^2$.
This interaction
features a dimensionless coupling constant in any
dimension. For our purposes it is sufficient to
consider two dimensions, where the coupling destroys conformal invariance.
As discussed above, the energy-momentum tensor
depends on the coupling to background curvature even in flat space.
It is




$$
T\equiv T_{zz} = - {1\over 2} (\partial\phi)^2 +\xi(~(\partial\phi)^2+
\phi\partial^2\phi)
$$
and the propagator is $\langle\phi(z)\phi(0)\rangle
= - {\rm log} z$ for all $\xi$. The $\xi$-dependent term is a total
derivative.
Under a conformal transformation
$z\rightarrow f(z)$ the energy--momentum tensor transforms as
$$
T(z)\rightarrow (f^{\prime}(z))^2 T(f(z)) + A_f(z)
$$
For $\xi=0$
the extension term
$$
A_f(z)^{\xi=0}\equiv {1\over 12}S_f (z) =
{1\over 12} {f^{\prime\prime\prime}f^\prime
-{3\over 2}(f^{\prime\prime})^2\over (f^\prime)^2}
$$
satisfies the associativity property
$$
A_y(z) = (x^\prime)^2 A_y(x) + A_x(z)~;~~z=z(x)
$$
Then succesive transformations on $T$ gives the same transformed
tensor as one combined transformation, {\it i.e.} the energy--momentum
tensor realizes the conformal group. In fact, the form of $A_f(z)$ is
determined by this property.
In the general case of non-minimal coupling, $\xi\neq 0$, we expect
conformal invariance to be broken, and $A_f(z)$ to be of a more
general form.

To calculate the extension term
we proceed as in [\cardy]. Noting that the formal expression
contains singular products of operators at the same point
we regulate by defining
$$
(\partial\phi)^2 \equiv [(\partial\phi(z+{1\over 2}d)\partial\phi
(z-{1\over 2}d)+{1\over d^2}]_{d\rightarrow 0}
$$
$$
\phi\partial^2\phi
 \equiv [\phi(z+{1\over 2}d)\partial^2\phi
(z-{1\over 2}d)-{1\over d^2}]_{d\rightarrow 0}
$$
The regulator $d$ is kept fixed under a conformal transformation.
After the transformation, a new
regulator $d^\prime=f(z+{1\over 2}d)-f(z-{1\over 2}d)$
is appropriate, however.
It is the difference of the two regulators that is the subtle origin of
the extension term. Explicitly,
$$
A_f(z)^{\xi=0} = {1\over 2}
[{f^\prime(z+{1\over 2}d)f^\prime(z-{1\over 2}d)\over
(f(z+{1\over 2}d)-f(z-{1\over 2}d))^2} - {1\over d^2}]_{d\rightarrow 0}
=
{1\over 12} {f^{\prime\prime\prime}f^\prime
-{3\over 2}(f^{\prime\prime})^2\over (f^\prime)^2}
$$
Similar treatment of $\phi\partial^2\phi$ leads to
$$
A_f(z)=(1-2\xi)A_f(z)^{\xi=0}+
\xi[{f^{\prime\prime}(z-{1\over 2}d)\over f(z+{1\over 2}d)-f(z-{1\over 2}d)}
+{f^\prime(z-{1\over 2}d)^2\over (f(z+{1\over 2}d)-f(z-{1\over 2}d))^2}
-{1\over d^2}]_{d\rightarrow 0}
$$ $$
={1\over 12} (1-6\xi) S_f(z) - {\xi\over 4}
({f^{\prime\prime}\over f^\prime})^2
$$
after a small calculation.
The extra $({f^{\prime\prime}\over f^\prime})^2$ term
violates the associativity property of $S_f(z)$ so
conformal invariance is not realized for $\xi\neq 0$.
This is no surprise, since the energy--momentum tensor derives from
a curved space action that explicitly breaks conformal invariance.

As the extension term appears as a consequence of the definition of
the quantum operator products, it is usually an anomaly which is not
present in the classical theory. Here,
however, the operator
$(\partial\phi)^2+\phi\partial^2\phi $ is regular a short distances
even though the individual terms are not. Therefore,
the $\xi$ dependent terms are not really anomalies, but rather
explicit breaking of conformal invariance. The complete extension
term is the sum of the anomalous ($\xi$ independent)
and explicit ($\xi$-dependent) contributions.

{}From the extension term we can calculate the constant term in the heat
kernel.
It is related, through the effective action, to the finite part of
the trace of the energy--momentum tensor
$$
D= -{1\over 2\pi}\int \langle T^\mu_\mu\rangle d^2 r =
- (\int z\langle T(z)\rangle {dz\over 2\pi i} + h.c.)
$$
If we insist on $\langle T(z)\rangle = 0 $ in flat space, then
$\langle T(z)\rangle = A_f(z)$ on the cone, where $f=z^{1\over n}$ maps
flat space on to a cone. Noting that in these coordinates $z$ traverses
an angle $2\pi n$ around the singularity, we find
$$
D = {1\over 12} ({1\over n} - n) - \xi (1-n)
$$
as derived heuristically above. While the
$({f^{\prime\prime}\over f^\prime})^2$-term is proportional to $(1-n)^2$,
so that it does not contribute to the entropy, it is exactly
such as to change the $\xi$-dependent
term from the non-perturbative $({1\over n}-n)$-form to the
perturbative $(1-n)$-form. In this calculation, performed without
any reference to curved space, we see how the total derivative
in the energy--momentum tensor enters the final result.


To round out the discussion, we will now briefly
analyze another possible form of nonminimal coupling.
Consider the Lagrangean
$$
{\cal L} = {1\over 8\pi} [ (\nabla\phi)^2 + 2\gamma R\phi ]~.
\eqn\othernm
$$
The parameter $\gamma$ is dimensionless in two dimensions and has the
dimension of mass in four dimensions. In flat space the $\gamma R\phi$
term vanishes but
the energy--momentum tensor
$$
T_{\mu\nu} = - {4\pi\over\sqrt{g}}
{\delta\int{\cal L}\sqrt{g}\over\delta g^{\mu\nu}} =
-{1\over 2}\partial_\mu\phi\nabla_\nu\phi +
{1\over 4}\eta_{\mu\nu}\nabla_\alpha\phi\nabla^\alpha\phi +
\gamma (\eta_{\mu\nu}\nabla^2-\nabla_\mu\nabla_\nu)\phi
$$
acquires a surface term.

In two dimensions, the holomorphic part of the energy--momentum
tensor is $T= -{1\over 2}(\partial\phi)^2 - \gamma\partial^2 \phi$.
We use the propagator $\langle\phi(z)\phi(0)\rangle=-{\rm log}~z$
to calculate the operator product expansion
$$
T(z)T(0) = {c\over 2z^4} + {2T(0)\over z^2} + {\partial T\over z}
+{\rm regular}
$$
where $c=1+12\gamma^2$. The form of the operator product expansion
shows that the energy--momentum tensor defines a conformal field theory.
This is by no means trivial. For example, the non-minimal coupling
$\xi R\phi^2$ would give rise to a term $\xi{{\rm log} z \over z^4}$
so that theory is too singular to realize conformal symmetry.
For conformal field theories in two dimensions
the geometric entropy is proportional to $c$ [\hlw ]. Thus
a $\gamma R\phi$ interaction, and  the ensuing surface term in the
energy--momentum tensor, directly affect
the geometric entropy in two dimensions.

In higher dimensions we proceed differently. Because the
the non-minimal term is linear in the field $\phi$, rather than quadratic,
we can not use the heat kernel to reduce the problem
effectively to two dimensions, as we did before.
Instead, we perform a Gaussian integral to
compute  the effective action
$$
W_{\rm div}(\gamma) - W_{\rm div}(\gamma=0) = {\gamma^2 \over 8\pi}
\int dx dx^\prime R(x)G(x,x^\prime)R(x^\prime)
$$
where $G={1\over \nabla^2}$ is the propagator.
To facilitate comparison with previous results we need a local form
of this result. We readily calculate
$$
(T_\mu^\mu)_{\rm div} = - 4\pi g^{\mu\nu}{\delta W_{\rm div}\over
\delta g^{\mu\nu}} =
(d-1)\gamma^2 R + (T_\mu^\mu)_{\rm div}^{\gamma=0}
= - 4\pi{\partial W_{\rm div}\over \partial{\rm ln}{1\over\epsilon^2}}
\eqn\tensvar
$$
In two dimensions,
we use the heat kernel result for the $\gamma=0$ case to find
$(T_\mu^\mu)_{\rm div}^{\gamma=0}={1\over 12}R$, and
thereby recover the anomaly calculated in conformal theory.
In four dimensions the $\gamma^2$-term also contributes a
logarithmic term to the effective action.
However the leading divergence in the renormalization
of Newton's constant is quadratic; thus this particular form
of non-minimal coupling does not contribute to the leading divergence.
A closer analysis shows that in fact it contributes only a finite
renormalization.

\section{Geometric Entropy and Fermions}

In this section and the next, we calculate the geometric entropy
contributed by matter fields with spin.  As we shall see, these
calculations rapidly reduce to the calculations already done for
non-minimally coupled scalars, with particular choices for the
parameter $\xi$.

The geometric entropy of a number of noninteracting species of particles
is equal to the entropy of each species by itself.
For fermions the contribution is positive,
as shown by an explicit calculation in [\fermions].
However, since loops of virtual bosons and fermions differ by an explicit
sign, reflecting the quantum statistics, one
might
expect that spin-0 and spin-1/2
contribute with opposite sign to the renormalization
of the gravitational coupling.
However upon closer scrutiny this heuristic argument
is incomplete -- and indeed, its conclusion is false --
due to the fact that for spin-1/2 one has an
effective non-minimal coupling to curvature.

Indeed consider the effective
action
$$
e^{-W} = \int {\cal D}\psi{\cal D}{\bar\psi}~
e^{-\int {\bar\psi} i\gamma^{\mu} D_\mu \psi}
={\rm det}~(i\gamma^\mu D_\mu)
=[{\rm det}~( -\gamma^\mu D_\mu \gamma^\nu D_\nu) ]^{1\over 2}~.
\eqn\feffs
$$
The positive power of the functional determinant
here, as contrasted with the
negative power in \bosoneff\ , reflects the
aforementioned sign.
In a frame where the connection $\Gamma =0$, we have
$$
 -\gamma^\mu D_\mu \gamma^\nu D_\nu =
  -\gamma^\mu\partial_\mu \gamma^\nu\partial_\nu-
{i\over 2}\gamma^\mu\gamma^\nu\partial_\mu
\Gamma^{\alpha~\beta}_{~\nu}\Sigma_{\alpha\beta}~~;~~~
\Sigma_{\alpha\beta}
={i\over 4} [\gamma_\alpha , \gamma_\beta ]~.
\eqn\gamiden
$$
To extract the most divergent contribution,
it suffices to take the Dirac trace at this early stage, and
one
finds
$$
e^{-W} = {\rm det} [-\Delta +{1\over 4}R]^{s\over 2}
\eqn\effcurvact
$$
where $s$ is the dimension of fermion representation, {\it e.g.}
$s=2$ for a Weyl fermion in four dimensions.
Using now
the heat kernel corresponding to
non--minimally coupled scalars, we
find
$$
\eqalign{
W({\rm fermion}) &= - s W({\rm boson}, \xi ={1\over 4})\cr
&= -s {{1\over 4}-{1\over 6}\over -{1\over 6}}W({\rm boson}, \xi=0)\cr
&={s\over 2}W({\rm boson}, \xi =0) ~.\cr}
\eqn\fermres
$$
Thus each spin-1/2 fermionic degree of freedom  contributes half as much
to the renormalization of
the gravitational coupling constant as a spin-0 bosonic one.

Nothing in the calculation prevents us from repreating it in a conical
background, finding
the same relation for the geometric entropies.
In two dimensions this follows from relation
between the corresponding conformal anomalies [\hlw ].
In heat kernel regularization the transverse dimensions
decouple; thus the result holds in four dimensions as well.
We see that for spin 1/2, as for spin 0, the divergences
in the gravitational coupling agree with those of the geometric
entropy in sign as well as in magnitude.


\section{Gauge Fields}


Now let us consider gauge fields. We write
$$
e^{-W} = \int {\cal D}A~e^{-{1\over 16\pi}\int F^2 }
= \int {\cal D}A~e^{-{1\over 8\pi}\int A_\mu (-g^{\mu\nu} \Delta
+\nabla^\mu \nabla^\nu + R^{\mu\nu} ) A_{\nu} }
$$ $$
= [{\rm det} (-g^{\mu\nu}\Delta + R^{\mu\nu} )]^{-{1\over 2}}_{V}
{\rm det} (-\Delta)_{S}
\eqn\gaugedet
$$
The final equality indicates we are
working in
Feynman gauge, for which
the gauge fixing term cancels the $\nabla^\mu\nabla^\nu$
term.  The scalar determinant derives from
the Jacobian involved in the gauge fixing,
which can be expressed in terms of a Fadeev-Popov ghost
field, {\it i.e.} a complex spin-0 field which is quantized as a fermion.
A careful derivation of this result, in a more general
context, can be found in [\vassilevich ].

To the accuracy we need, take the trace over the vector representation
(V) first and find
$$
e^{-W} = [{\rm det}(-\Delta + {1\over d}R)]_S^{-{d\over 2}}~{\rm det}
(-\Delta)_S
$$
{\it i.e.} the different polarizations decouple.
We immediately obtain
$$
W_{\rm gauge~field} = d~W_{\rm boson} (\xi = {1\over d})
- 2W_{\rm boson} (\xi=0) ~.
\eqn\gaugeres
$$

The preceding calculations apply equally to
the conical backgounds relevant to geometric entropy,
and to the almost flat smooth background relevant to renormalization
of the Newton constant.
Thus we conclude for a spin-1 gauge field, as for
spin 0 and spin-$1\over 2$, the gravitational
coupling constant and the geometric entropy have equal divergences.
Quantitatively
$$
{1\over G_{\rm ren}\hbar} = {1\over G_{\rm bare}\hbar} + {1\over 2\pi}
({(d-2)\over 6}-1) {1\over\epsilon^2}~.
\eqn\renres
$$

For $d=4$ this agrees with Kabat [\kabattwo], who performed a
very careful and explicit mode analysis, and earlier work [\birrell ].
Our derivation appears to highlight the basic similarity of
different spins.
In particular, gauge invariance specifies a unique
non-minimal coupling
to the background curvature, but plays no other role.
This unique coupling leads to a negative value for the
leading divergent contribution to the covariantly regulated
geometric entropy.
This is at first sight a startling phenomenon, but as we
have seen it can occur already, and for similar reasons, in the
context of
non-minimally coupled scalars.




\chapter{Discussion}

$\bullet$ For the preceding analysis it has been crucial to use
a local, covariant regulator
throughout.  The heat kernel method supplies a convenient
regulator of this source, and in addition
leads to very simple calculations,
as we have seen.  A pioneering analysis of divergences
in the entropy [\thooftone], which (with minor variations)
has been widely
adopted, employed instead a physically motivated
scheme in which one first
evaluates local entropy density, and then integrates over space
to find the total entropy. The divergence of the latter integral
expresses the divergence in local temperature close to the horizon.
A regulating cutoff is imposed at some specified physical
distance from the horizon.  Clearly there is considerable
arbitrariness
in the choice of distance, and it is difficult to compare this
regulation of the black hole entropy to any convenient regulator
for quantum fluctuations in smooth geometries.
While the choice of regulator scheme
cannot ultimately affect physical results, an unfortunate choice may
obscure the physics by requiring complicated conspiracies among
counterterms.  Our choice has the virtue of allowing
straightforward comparison between the
regulated Newton's constant and Gibbons-Hawking or geometric
entropies,
as we have seen. Similar issues were recently emphasized in [\demers ].

$\bullet$  In calculating the entropy of
a large black hole it seems natural
to include the classical entropy
known from black hole thermodynamics together with the
geometric entropy of quantum origin.
However, this procedure is somewhat {\it ad hoc}, and
it is not entirely clear how it should be formulated
for a finite black
hole, or even for pure vacuum (recall that in the Gibbons-Hawking
calculation the divergent flat-space piece was simply subtracted off).

Alternatively, one could entertain the hypothesis
that gravity itself
induces a natural cut-off which makes geometric entropy, without
the addition of a separate classical term, take on the finite value
${A\over 4G_R{\hbar}}$.  The equality of
entropy and coupling renormalization discussed here,
if it still applies, would then seem to indicate
the renormalized gravitational coupling constant arises
entirely from quantum corrections.
This, of course, is the basic hypothesis of induced
gravity [\adler ].

How might such a cut-off conceivably arise?
The gravitational coupling apparently
suffers no ultraviolet divergences
in string theory [\gsw ]. The connection discussed in this paper
suggests that black hole entropy and
the geometric entropy must likewise be finite in string theory.
Unfortunately our understanding of large
black holes and of  space-times
with boundary in string theory remains primitive, and it appears
very challenging at present
to substantiate these suggestions, or even to make
them completely precise.

An alternative possibility, which does not necessarily contradict
the previous one,
is that properly implementing the constraints of gravity
drastically reduces the number of states, and itself renders
the various entropies finite.
Bekenstein has forcefully advocated this
possibility on a variety of physical grounds [\bekenstein ].
It appears amenable to investigation
at the semiclassical level [\perthree ].

{\bf acknowledgements}

We wish to thank P. Kraus and A. Peet for stimulating discussions.

\refout

\end